# Range-Doppler-Acceleration Estimation for Fast-Moving and Accelerating Targets

Nadav Neuberger | Simon Kollecker | Martin Käske

Fraunhofer Institute for High Frequency Physics and Radar Techniques FHR, Wachtberg, Germany

**Correspondence:** Nadav Neuberger (nadav.neuberger@fhr.fraunhofer.de)

## ABSTRACT

Fast-moving and/or accelerating targets challenge conventional pulsed-radar processing because the received echo may exhibit both pulse-to-pulse migration and intra-pulse motion-induced distortion. In such cases, standard range–Doppler processing based on stop-and-go or negligible intra-pulse motion assumptions can suffer substantial loss of focus and estimation accuracy. This paper presents a generalised framework for range–Doppler–acceleration estimation that remains compatible with efficient FFT-based processing while accounting for these effects. The approach is based on a Cruise-and-Go approximation, which linearises the target motion within each pulse, leading to both carrier frequency Doppler shift and baseband signal stretch, while allowing pulse-to-pulse variation. The resulting method is waveform independent and applicable to arbitrary frequency- and phase-coded signals. Its performance is analysed in terms of estimation accuracy, computational complexity, and ambiguity behaviour, with comparisons to existing methods. A unified metric is introduced to predict correlation loss. Simulations show accurate refocusing and estimation performance close to the Cramér–Rao bound, making the method suitable for long-pulse and wideband radar applications, including space surveillance.

## 1 | Introduction

In radar Signal Processing (SP), the handling of Fast-Moving and/or Accelerating Targets (FMAT) is of central importance. The received echo from such a target is not only delayed relative to the transmitted waveform but also distorted by Doppler effects, including frequency shifts and time scaling. When multiple pulses are coherently integrated within a Coherent Processing Interval (CPI), these motion-induced effects lead to range and/or Doppler cell migration, thereby degrading the output Signal-to-Noise-Ratio (SNR) and the accuracy of parameter estimation, if not properly taken into account in the processing of Range-Doppler (RD) estimation.

The severity of these distortions depends jointly on the target's kinematic behaviour (e.g., speed, acceleration, and degree of motion non-linearity) and on system parameters (e.g., bandwidth and CPI duration). Three practical regimes can be distinguished. (1) Slow motion: The target remains within a single range cell and the radial motion can be treated as approximately constant across the CPI. Standard two-dimensional Fast Fourier Transform (FFT) processing is therefore sufficient [1]. (2) Linear but significant motion: The motion is still linear over the CPI but large enough to cause range migration. The Keystone Transform (KT) is widely used for compensating this effect through slow-time rescaling in a Linear Frequency Modulation (LFM) waveform [2]. (3) High and/or non-linear motion: When the radial velocity is high, the waveform experiences intra-pulse Doppler shifts and time stretching; when the range evolution is quadratic, energy spreads across both range and Doppler dimensions. In such cases, 2D-FFT processing and KT become inadequate, since the waveform distortion must be handled on a per-pulse basis [3]. As an example, scenarios of this type arise in Space Situational

Awareness (SSA), where satellites may exhibit radial accelerations on the order of $200 \mathrm{ms}^{-2}$, and in long-pulse or wideband radar systems where intra-pulse Doppler cannot be ignored. While it is possible to mitigate intra-pulse effects due to high Doppler by decomposing the signal into smaller parts, each of which is Doppler-agnostic, this only makes sense if the range domain of interest is very small; otherwise there will be a large overlap between partitions.

There are multiple ways in the literature to deal with FMAT, such as the Discrete Polynomial-Phase Transform (DPT) [4, 5]. It estimates acceleration by modelling the slow-time signal as a polynomial phase sequence, where higher-order motion terms are converted into lower-order phase components through phase differencing and estimation via Fourier analysis. The Radon-based methods, such as [6] and Radon-Fractional Fourier Transform (RFRFT) [7], perform coherent integration along hypothesised motion trajectories in the range–slow-time plane, often combining Radon search with fractional Fourier or Doppler filtering to compensate for range migration and Doppler frequency modulation. Such methods explicitly search over motion parameters and can effectively refocus manoeuvring targets under constant-acceleration models. KT-based methods with higher-order extensions [8–10] attempt to mitigate quadratic migration terms. While computationally efficient and widely adopted, they rely on the *Stop-and-Go* (SnG) assumption.

Importantly, most of these methods ignore intra-pulse motion, and all of them ignore intra-pulse baseband stretch. While waveform-independent in formulation, the vast majority of implementations in the literature refer to LFM. Its associated RD coupling and Doppler ambiguity issue is addressed under restrictive scenario conditions in [11]. Thus, excluding relevant waveform modulations such as Noise-Radar [12], Costas-Code [13], diverse pulses train [14], phase-modulation or amplitude weighting.

This work extends the approach introduced in [15], and offers novel contributions with respect to Range-Doppler-Acceleration (RDA) estimation in the FMAT regime, by compensating for both inter-pulse acceleration *and* intra-pulse motion-induced time distortion. Furthermore, an accurate delay expression is employed to provide a clear view of the approximations used. It is applicable and formalised to *any* frequency or phase coded waveform. Within this process, we coin a new approximation *Cruise-and-Go* (CnG) that linearises the delay within each pulse while allowing pulse-to-pulse dynamics, enabling FFT-based matched filtering for quadratic motion. We use a unified performance metric that predicts the framework SNR loss and accuracy, along with the method's operational limits. An important analysis of Doppler-ambiguity behaviour in the FMAT regime is discussed with recommendations for waveform selection.

The remainder of this paper is organised as follows. Section 2 introduces the signal model, including the time-delay formulation under quadratic radial motion. Section 3 reviews the classical RD processing chain and highlights the assumptions that break down for FMAT. Section 4 presents the proposed method, including the CnG approximation and its resulting FFT-based implementation. Section 5 evaluates the method through simulations, including motion compensation, intra-pulse stretch effects, and Doppler ambiguity behaviour. In Section 6 we quantify its limitations, introduce a unified performance metric for predicting SNR loss and statistical noise analysis. Section 7 concludes the paper and summarises the main findings. The Appendix provides an analytical assessment of the approximation error introduced by the CnG assumption.

## 2 | Signal Model

In this section, we describe a pulsed radar signal model, taking into account both radial velocity and acceleration. In some radar applications (e.g., SSA), the potential targets could be thousands of kilometres away, which requires the use of unambiguous long Tx pulses, and high energy transmission. In addition, the radial velocity and acceleration can reach high values, and the Doppler shift can have an impact even within a single pulse.

We consider a single moving target, within a single CPI, for a single noise-free antenna. We denote the following parameters: the single CPI consists of $N_\mathrm{p}$ identical pulses. Its Pulse Repetition Interval (PRI) is $T_\mathrm{pr}$ and CPI is $T_\mathrm{cp} = T_\mathrm{pr} N_\mathrm{p}$. The fast time variable is $\Delta t$, while $T_m$ is the slow-time variable, depending on the $m$-th pulse index $T_m = T_\mathrm{pr}(m-1)$, where $0 \leq \Delta t \leq T_\mathrm{pr}$. We emphasise that $t \in [t_0, t_0 + T_\mathrm{cp}]$ represents the elapsed time since the CPI start. For simplicity, we choose $t_0 = 0$ since we will consider a single CPI throughout the paper.

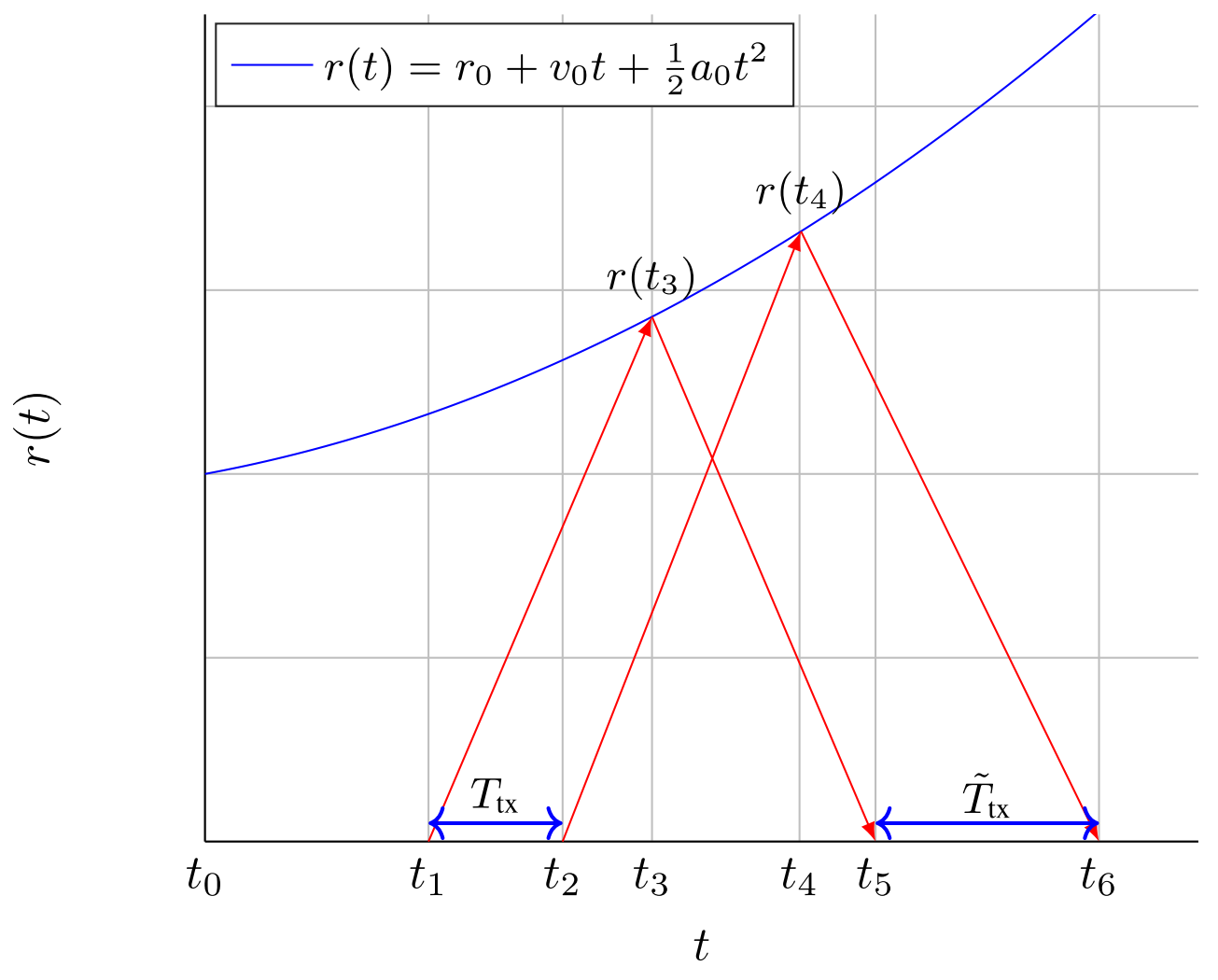

**FIGURE 1** | Delay induced on transmitted signal by a target with quadratic range-profile. Within one pulse, we illustrate two transmitted samples (beginning and end of the waveform) at times $t_1, t_2$, and received at times $t_5, t_6$ respectively. The stretching effect is shown due to the high radial velocity–the received pulse length $\tilde{T}_\mathrm{tx}$ is longer than $T_\mathrm{tx}$.

## 2.1 | Time-Delay

We start with a single pulse derivation, and extend it later to a train of pulses. In Figure 1 we illustrate the range of a single target (i.e., distance between Tx/Rx and target) having a quadratic range-profile. Note that the scenario is exaggerated for illustrative purposes especially with respect to the speed-of-light $c_0$ and thus the time it takes the pulse to intercept the target. We distinguish the following parameters: $t$ is the global timestamp (in seconds), elapsed time since the beginning of the CPI, named $t_0$. Let $\tau(t)$ be the two-way delay of a sample transmitted at $t_{tx}(t)$, reflected at $t'$, and received at time $t$. This delay depends on the target range at $t' = t - \tau(t)/2$.

The transmission time stamp $t_{tx}(t)$ corresponds to an arbitrary sample of the Tx pulse. For example, for a signal received at time $t_5$, we have $t' = t_3$ and $t_{tx}(t_5) = t_1$, and $\tau(t_5) = t_5 - t_1$. The stretching effect is shown due to the high radial velocity, as the received pulse length is longer than the transmitted one. We can now formulate

$$\tau(t) = t - t_{tx}(t) = \frac{2}{c_0} r(t - \tau(t)/2). \tag{1}$$

This accurate implicit expression of the delay is commonly replaced with $\tau(t) = 2r(t)/c_0$, which approximates the range of the target at time of interception as $t$ instead of the accurate value $t'$.

The down-converted baseband signal echo that arrives to a single antenna at time snapshot $t$, for a single pulse, caused by a single point target with parameters $\theta_{tg}$ can be expressed as

$$x(t, \theta_{tg}) = A \underbrace{s[t - \tau(t, \theta_{tg})] e^{-i2\pi f_c \tau(t, \theta_{tg})}}_{b(t, \theta_{tg})}. \tag{2}$$

$\tau$ is the fast-time dependent two-way delay of a wave reflected by the target and arriving at the receiver at times $t$. It depends on the target's parameters vector $\theta_{tg}$ at the initial timestamp of the CPI $t_0$. The carrier wave frequency is $f_c$, and the amplitude $A$ depends on the range (representing the propagation loss), the target's Radar Cross Section (RCS), and antenna Direction of Arrival (DOA) two-way gain. For simplicity and without loss of generality, we will assume it to be unity and ignore any DOA aspects for the remainder of the text.

Equation (2) shows two dependencies of the signal on $\tau$, both in the argument of $s$ and in the phase term in the end. The effect of a linear change in $\tau$ on the phase term is typically referred to as a Doppler shift, whereas its effect on the signal part in the beginning is called range migration. Typically, these effects are only observed from pulse to pulse, but this work will also consider them inside the pulse. They will be referred to as intra-pulse Doppler shift and intra-pulse baseband stretch respectively.

We will treat the case of a quadratic time-dependent range, where $\theta = [r_0, v_0, a_0]$, with $r_0, v_0, a_0$ as the initial range, radial velocity and acceleration of the target, respectively. The range follows

$$r(t, \theta) = r_0 + v_0 t + \frac{1}{2} a_0 t^2. \tag{3}$$

From Equations (1 and 3), we can now write

$$\frac{c_0}{2} \tau(t) = r_0 + v_0 (t - \tau(t)/2) + \frac{1}{2} a_0 (t - \tau(t)/2)^2. \tag{4}$$

An analytic expression for $\tau(t)$ can be found by first considering $\tau$ and $t$ to be independent variables and finding the roots of the quadratic function w.r.t. $\tau(t)$. However, the solution is highly nonlinear which prohibits further efficient computation methods. In the next sections we will show two different approaches to approximate $\tau(t)$.

## 2.2 | Range-Doppler Compression

One central aspect in every radar SP scheme is the RD compression. This crucial and computationally demanding process estimates the RD values of the target, under an assumed motion model. The unknowns in $\theta$ are estimated using a correlation (or Matched Filter (MF)) based approach where the actual received data is correlated with the template signal. The set of parameters maximising the correlator (absolute) output is used as an estimate of $\theta_{tg}$. We now write the RD equation of a single Rx pulse, for the specific target's parameters $\theta_{tg}$, and an estimated target's parameters $\theta$ as

$$\begin{aligned} \mathrm{RD}(\theta) &= \int x(t, \theta_{tg}) b^*(t, \theta) dt \\ &= \int x(t, \theta_{tg}) s^*[t - \tau(t, \theta)] e^{i2\pi f_c \tau(t, \theta)} dt, \end{aligned} \tag{5}$$

where $x(t, \theta_{tg})$ is the received measured baseband signal, and $b$ is used as the template signal. The SNR is derived from the response magnitude and the known statistical noise distribution. When integrating a series of $N_p$ pulses, the complex (coherent) or magnitude (non-coherent) output of Equation (5) is summed up for all pulses. Its direct time-domain calculation has a heavy computational load, which prevents any practical response time.

# 3 | Classical Method

There are several simplifications and assumptions that can be made to compute Equation (5) using FFT. These are very common, usually permitted by the application scenario and allow a fast and practical computation time. We present this simplified scenario derivation as a theoretical basis and motivation for our new method in Section 4.

In the case of a single pulse, where the radial velocity of the target is low such that the motion within the pulse is negligible in terms of range resolution and phase, we can simplify Equation (1) into

$$\tau(t) \approx \frac{2r(t_0, \theta)}{c_0} = \frac{2r_0}{c_0} = \tau_0, \tag{6}$$

and re-write Equation (5) as

$$\mathrm{RD}(\theta) = e^{i2\pi f_c \tau_0} \int x(t, \theta_{tg}) s^*[t - \tau_0] dt.$$

The next step is crucial: transforming the equation into two separate functions, to allow a time-convolution equivalent with an FFT operation. The definition of a convolution is

$$(u(t) \circledast v(t))(\psi) = \int u(t) v(\psi - t) dt. \tag{7}$$

We denote two auxiliary functions

$$u(t) = x(t), \quad v(t) = s^*(-t) \tag{8}$$

such that

$$\mathrm{RD}(\theta) = e^{i2\pi f_c \tau_0} (u(t) \circledast v(t))\big|_{\psi = \tau_0}. \tag{9}$$

The complex output value represents the magnitude and phase of the target echo. Since time-convolution is equivalent to multiplication in the Fourier domain, we can write

$$\mathrm{RD}(\tau_0) = e^{i2\pi f_c \tau_0} \mathcal{F}^{-1}\{X(f) V(f)\}\big|_{t = \tau_0}, \tag{10}$$

where $X(f) = \mathcal{F}(u(t))$, $V(f) = \mathcal{F}(v(t))$, and $\mathcal{F}$ is the FFT operator. This step is the so-called 1D *Range Compression* which is done as a first stage in the SP sequence. We note that $V(f)$ can be pre-calculated once ('offline'), leaving only the calculation of $X(f)$ in real-time. For a pre-defined set of values of $\tau_0$, we choose the maximum absolute value for Equation (10) as the estimate of the target's delay (or range)

$$\hat{\tau}_0 = \underset{\tau_0}{\operatorname{argmax}} |\mathrm{RD}(\tau_0)|. \tag{11}$$

When integrating multiple pulses, the SnG approximation is widely used. It assumes the target does not move during the transmission of a single pulse; its location only updates once between one pulse and the next one. Hence, the RD output of each pulse will depend on the pulse index.

With the assumption of a constant radial velocity throughout the CPI, the two-way target delay at the start of the $m$-th pulse is

$$\tau_m = \tau_0 + \frac{2 v_0 T_m}{c_0}, \tag{12}$$

where the '0' subscript points to the beginning of the CPI. The SnG assumption states that $\tau(t) = \tau_m$ for $T_m < t < T_{m+1}$. By plugging this expression in Equation (5) we get

$$\begin{aligned} &RD_m(\tau_0, v_0) = \\ &e^{i2\pi f_c \frac{2 v_0 T_m}{c_0}} e^{i2\pi f_c \tau_0} \int x_m(t, \theta_{tg}) s^*\left[t - \tau_0 - \frac{2 v_0 T_m}{c_0}\right] dt. \end{aligned} \tag{13}$$

By disregarding any range migration due to the low velocity, poor range resolution, or short CPI, we reach the coherent integration expression

$$\begin{aligned} &\mathrm{RD}(\tau_0, v_0) = \\ &e^{i2\pi f_c \tau_0} \sum_{m=1}^{N_\mathrm{p}} e^{i2\pi f_c \frac{2 v_0 T_m}{c_0}} \int x_m(t, \theta_{tg}) s^*[t - \tau_0] dt. \end{aligned} \tag{14}$$

This could be seen as a two-step process, first the complex output FFT range compression given by Equation (10), followed by another Inverse Fast Fourier Transform (IFFT) over slow-time with the Doppler frequency shift $f_d = -2 v_0 f_c / c_0$. Taking the magnitude gives

$$\begin{aligned} |\mathrm{RD}(\tau_0, v_0)| &= \\ &\left| \mathcal{F}_{\mathrm{st}}^{-1} \left\{ \mathcal{F}_{\mathrm{ft}}^{-1}\{X_m(f) V(f)\}\big|_{t = \tau_0} \right\} \Big|_{f = f_d} \right|, \end{aligned} \tag{15}$$

where the subscripts *st* and *ft* mean slow time and fast time inverse Fourier transform. The values of $\tau_0, v_0$ that maximise Equation (15) are the RD estimations. Using two sequential IFFTs provides a fast output, and as we see next, cannot be used when some of the simplifications above are prohibited. However, the notion of exploiting FFT is the main drive for our work.

## 4 | CnG RDA Estimation Method

Although the classical method provides computational efficiency, it fails under FMAT conditions. The motivation for the proposed CnG-based framework is therefore threefold. First, it compensates for interpulse radial acceleration effects, including both phase accumulation and migration in range-velocity space. Second, it accounts for intra-pulse radial velocity variations without invoking the SnG assumption, thereby correcting both intra-pulse Doppler shift and intra-pulse range migration, which is also called baseband stretch. Third, it maintains waveform independence, extending applicability beyond LFM signals. This generalised approach is realised through a single linearisation step while retaining the computational advantages of FFT-based fast-time processing.

### 4.1 | CnG Approximation

While the true motion is quadratic, the velocity change within a single pulse is usually small compared to its change across the CPI. Therefore, we approximate the velocity within each pulse as constant, but allow it to vary per pulse. This linearises the delay $\tau(t)$ with respect to $\Delta t$, enabling matched filtering via standard FFT operations with only frequency remapping. We coin the term *Cruise* for constant-velocity motion and *go* when acceleration resumes.

The time $t$ is formalised as a sum of fast and slow time terms, $t = T_m + \Delta t$. We denote $s(t)$ as the transmitted baseband signal (a single pulse with duration $T_\mathrm{p}$) as $s(t) = s(T_m + \Delta t) = s_m(\Delta t) = s(\Delta t)$. The range and radial velocity at the beginning of the $m$-th pulse are (see Equation (3))

$$\begin{aligned} v_m &= v_0 + a_0 T_m \\ r_m &= r_0 + v_0 T_m + \frac{1}{2} a_0 T_m^2. \end{aligned} \tag{16}$$

We now explicitly write the CnG assumption, where the radial velocity within each pulse is constant (i.e., $a_m = 0$) such that

$$\begin{aligned} v(T_m < t < T_{m+1}) &= v_m \\ r(T_m < t < T_{m+1}) &= r_m + v_m \Delta t. \end{aligned} \quad (17)$$

## 4.2 | Mathematical Derivation

Solving Equation (1) will now give

$$\tau(t) = \frac{2}{c_0} r(t - \tau(t)/2) = \frac{2}{c_0}(r_m + v_m \cdot (\Delta t - \tau(t)/2)),$$

which yields

$$\begin{aligned} \tau(t) &= \phi_m + \gamma_m \Delta t, \\ \phi_m &= \frac{2r_m}{c_0 + v_m},\ \gamma_m = \frac{2v_m}{c_0 + v_m}. \end{aligned} \quad (18)$$

The parameter $\gamma_m$ describes the intra-pulse Doppler-induced time scaling. Going back to the RDA expression, we plug Equation (18) into Equation (5) and derive the $m$-th pulse RDA output as

$$\begin{aligned} &\mathrm{RDA}_m(\theta) \\ &= \int x_m(\Delta t, \theta_{tg}) s^*[\Delta t - \tau(\Delta t, \theta)] e^{i2\pi f_c \tau(\Delta t, \theta)} d\Delta t \\ &= e^{i2\pi f_c \phi_m} \int \{x_m(\Delta t, \theta_{tg}) s^*[(1-\gamma_m)\Delta t - \phi_m] \\ &\qquad \times e^{i2\pi f_c \gamma_m \Delta t}\} d\Delta t \\ &= e^{i2\pi f_c \alpha_m} \int \{x_m(\Delta t, \theta_{tg}) s^*[-(1-\gamma_m)(\alpha_m - \Delta t)] \\ &\qquad \times e^{-i2\pi f_c \gamma_m (\alpha_m - \Delta t)}\} d\Delta t, \end{aligned} \quad (19)$$

with $\alpha_m = \phi_m / (1 - \gamma_m) = 2r_m / (c_0 - v_m)$.

Following the convolution-based formulation used in Section 3, we choose two auxiliary functions

$$\begin{aligned} u_m(t, \gamma_m) &= x_m(t) \\ v_m(t, \gamma_m) &= s^*(-(1-\gamma_m)t) e^{-i2\pi f_c \gamma_m t}, \end{aligned}$$

such that

$$\mathrm{RDA}_m(\theta) = e^{i2\pi f_c \alpha_m} (u_m(t) \circledast v(t, \gamma_m))(\alpha_m). \quad (20)$$

In the Fourier domain it yields

$$\begin{aligned} \mathrm{RDA}_m(\theta) &= \frac{1}{|1-\gamma_m|} e^{i2\pi f_c \alpha_m} \\ &\quad \times \mathcal{F}^{-1}\left\{ X_m(f) S^*\left(\frac{f + f_c \gamma_m}{1-\gamma_m}\right)\right\}\Big|_{\psi = \alpha_m}. \end{aligned} \quad (21)$$

We can express $\alpha_m$ in terms of scaled delay plus a motion-induced offset $\zeta_m$

$$\alpha_m = \frac{\phi_m}{1-\gamma_m} = \frac{2r_0}{c_0 - v_m} + \zeta_m,$$

where $\zeta_m = 2(v_0 T_m + 0.5 a_0 T_m^2)/(c_0 - v_m)$. In the Fourier domain, Equation (21) writes as

$$\begin{aligned} \mathrm{RDA}_m(\theta) &= \frac{1}{|1-\gamma_m|} e^{i2\pi f_c \alpha_m} \\ &\quad \times \mathcal{F}^{-1}\left\{ X_m(f) S^*\left(\frac{f + f_c \gamma_m}{1-\gamma_m}\right) e^{i2\pi f \zeta_m}\right\}\Big|_{\psi = \frac{2r_0}{c_0 - v_m}}. \end{aligned} \quad (22)$$

We now introduce the auxiliary parameter

$$\rho_m = (c_0 - v_m)/c_0, \quad (23)$$

which represents the required frequency-domain scaling. It allows the RDA output to be evaluated on a uniform delay grid without time-domain interpolation. The output can now be sampled at the final desired delay points $\psi = \tau_0 = 2r_0/c_0$:

$$\begin{aligned} \mathrm{RDA}_m(\theta) &= \frac{\rho_m}{|1-\gamma_m|} e^{i2\pi f_c \alpha_m} \\ &\quad \times \mathcal{F}^{-1}\left\{ X_m(\rho_m f) S^*\left(\rho_m \frac{f + f_c \gamma_m}{1-\gamma_m}\right) e^{i2\pi \rho_m \zeta_m f}\right\}\Big|_{\psi = \tau_0}. \end{aligned} \quad (24)$$

This way, we can pre-compute $S(f)$ and $X(f)$ with a large upsampling factor, and then use (linear) interpolation to get the exact query values. Coherent integration across all pulses yields

$$\mathrm{RDA}(\theta) = \sum_{m=1}^{N_p} \mathrm{RDA}_m(\theta). \quad (25)$$

## 4.3 | Process Flowchart

The overall processing chain is summarised in Figure 2. The left branch defines the set of radial velocity and radial acceleration hypotheses according to the expected scene, and the subsequent processing is repeated for each hypothesis pair. Given the radar parameters (slow-time sampling), the auxiliary parameters are first computed. The transmitted template and the received pulse train are then upsampled, the query points implied by Equation (24) are evaluated, and both spectra are interpolated on these points. Finally, the coherent summation in Equation (25) is performed, yielding a 3D RDA map on which detection and parameter estimation are carried out. We note that the number of detected targets is independent of the hypothesis set size $\mathcal{H}$.

## 5 | Performance Comparison

In this section, three representative scenarios are examined to demonstrate the proposed method's effectiveness. The first scenario addresses targets undergoing radial acceleration, necessitating pulse-to-pulse (slow-time) compensation. The second scenario considers targets that exhibit range cell migration during both the pulse duration and the electromagnetic wave propagation time, commonly referred to as intra-pulse baseband stretch. The third scenario investigates the Doppler ambiguity resolution with two different waveforms.

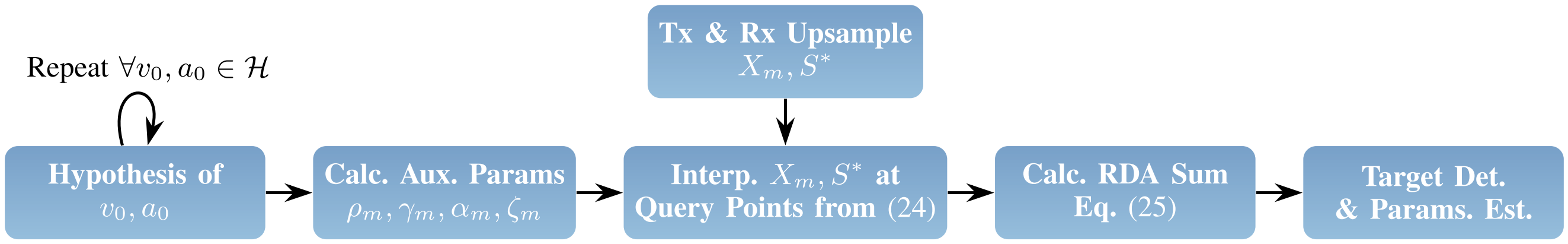

**FIGURE 2** | Process flowchart for a single CPI. Repeats for every feasible combination of $v_0, a_0$.

**TABLE 1** | Simulations parameters values.

| Symb. | Parameter | Sim. I | Sim. II | Sim. III |
|---|---|---|---|---|
| $f_c$ | Carrier frequency | 1.3 GHz | 2 GHz | 1.3 GHz |
| $B$ | Bandwidth | 8 MHz | 80 MHz | 2 MHz |
| $T_{pr}$ | PRI | 5 ms | 30 ms | 27 ms |
| $T_p$ | Pulse duration | 2 ms | 6 ms | 4.5 ms |
| $r_0$ | Range of target | 310 km | 1000 km | 1000 km |
| $v_0$ | Rad. vel. of target | 500 ms$^{-1}$ | 4000 ms$^{-1}$ | 0 ms$^{-1}$ |
| $a_0$ | Rad. acc. of target | 300 ms$^{-2}$ | 0 ms$^{-2}$ | 0 ms$^{-2}$ |
| $N_p$ | No. of pulses | 120 | 5 | 6 |

The various parameters used for these scenarios are summarised in Table 1. Unless stated otherwise, the simulations use a Costas waveform to isolate motion-induced distortions from waveform-induced RD coupling, enabling a clean evaluation of the processing method.

### 5.1 | Inter-Pulse Acceleration Compensation

As a basic motivation, we start with an example consisting of a long CPI, where compensation of radial acceleration is necessary to achieve a focused peak. We consider a scenario that induces pronounced nonlinear RD migration (see Sim. I from Table 1). Figure 3 shows two RD maps of a target with an initial radial velocity of 500 ms$^{-1}$ and a radial acceleration of 300 m s$^{-2}$, processed according to Equation (25). These scenarios are representative of spaceborne objects and airborne platforms. In the left panel, the data are processed under the incorrect assumption of zero acceleration ($a_0 = 0$).

The resulting response is severely defocused, mainly along the Doppler dimension-because the target velocity increases from 500 ms$^{-1}$ to approximately 680 ms$^{-1}$ over the 0.6 s CPI. This velocity change spreads the target energy across many Doppler bins and reduces the peak SNR by nearly 20 dB. Moreover, the defocusing follows a curved trajectory rather than a vertical ridge, which prevents coherent integration of energy from a single range cell, or any de-chirping methods.

The right plot shows the result when the correct acceleration ($a_0 = 300$ m s$^{-2}$) is applied. Here, the energy is sharply focused in both range and Doppler, confirming that acceleration compensation is essential for long CPIs involving targets with significant radial dynamics. This example primarily illustrates the need for *inter-pulse* acceleration compensation.

### 5.2 | Intra-Pulse Baseband Stretch

This aspect of comparison will illustrate the need for pulse-wise stretch compensation. When pulses are long and radial velocity is high, the frequently used SnG approximation is no longer valid (see Sim. II from Table 1). This comparison applies for any method that doesn't incorporate pulse-wise compensation.

Figure 4 shows the CnG processing, with and without pulse-wise baseband stretch compensation. In red, the factor $1 - \gamma_m$ in Equation (24) was used. We see an SNR degradation of 15 dB and a worst case range bias of 50 m (which spans over multiple range cells).

### 5.3 | Doppler Ambiguity Considerations

In the scenarios considered throughout the paper, the radial velocity usually exceeds the unambiguous Doppler interval defined by the Pulse Repetition Frequency (PRF), and the resulting ambiguities must therefore be resolved. It should be emphasised that these ambiguities do not originate from any processing method, but arise naturally in these scenarios. Several primary methods exist for resolving them, provided the radar parameters allow it.

One approach relies on range migration, which can resolve Doppler ambiguities when both the CPI and signal bandwidth are sufficiently large, for example [11]. The second approach is single-pulse Doppler processing, which requires a high duty cycle, adequate target SNR, and a Doppler-intolerant waveform. A third solution could be the use of staggered PRI, at the cost of dynamic range. To illustrate the issue and its solution using the second approach, Figure 5 shows the simulation of the parameters described in Sim. III from Table 1.

The RD maps are computed with CnG for a noise-free, stationary target. The first example uses a LFM waveform (Doppler-tolerant), while the second one uses a Costas code (Doppler-intolerant), both with a duty cycle of approximately 16 %. In this example, the low bandwidth and PRF, as well as the short CPI prevent the use of range migration for radial velocity ambiguity resolution. Since the PRF is constant, single pulse Doppler is the only remaining option.

In the LFM case, the output is inherently ambiguous, and the RD coupling is clearly visible: a shift in radial velocity can be compensated by a corresponding shift in range. The work in [11] proposes a solution based on range migration and a hypothesis filter bank spanning the Doppler ambiguous velocity space. In contrast, the Costas code produces only a few significant ambiguity sidelobes near the true peak, and only the correct Doppler bin attains full coherent gain. A sufficiently strong SNR is required in order to choose the main (correct) peak. The maxima of the first 4 ambiguities have an SNR of $-0.41$ dB, $-1.7$ dB, $-4.9$ dB, and $-7.9$ dB relative to the main peak. Those values mainly depend on the duty cycle $T_{\mathrm{p}}/T_{\mathrm{pr}}$ of the system, where a higher duty cycle corresponds to greater separation between the main lobe and ambiguity sidelobes.

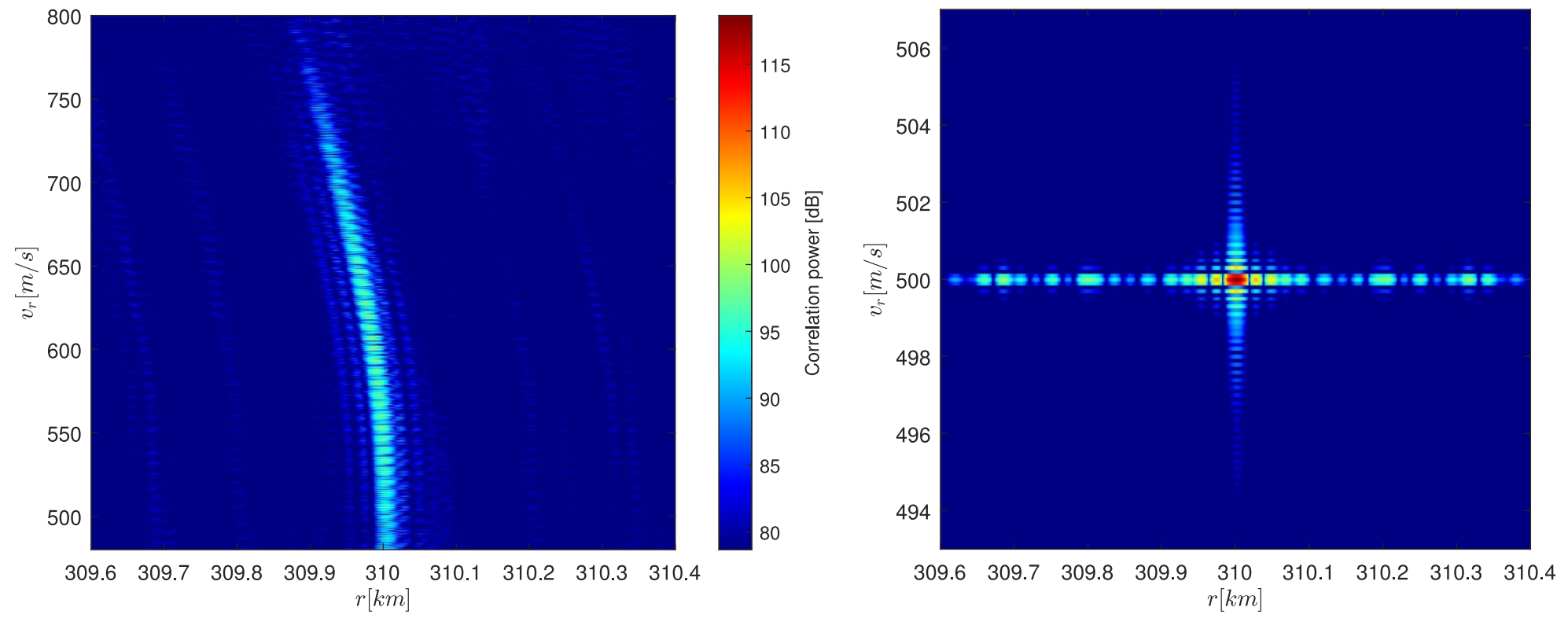

**FIGURE 3** | RD maps with and without acceleration compensation. The corresponding parameters are listed in (Table 1-Sim. I). The method is compensating for $a_0 = 0$ in the left plot and for the correct $a_0 = 300$ m s$^{-2}$ in the right plot.

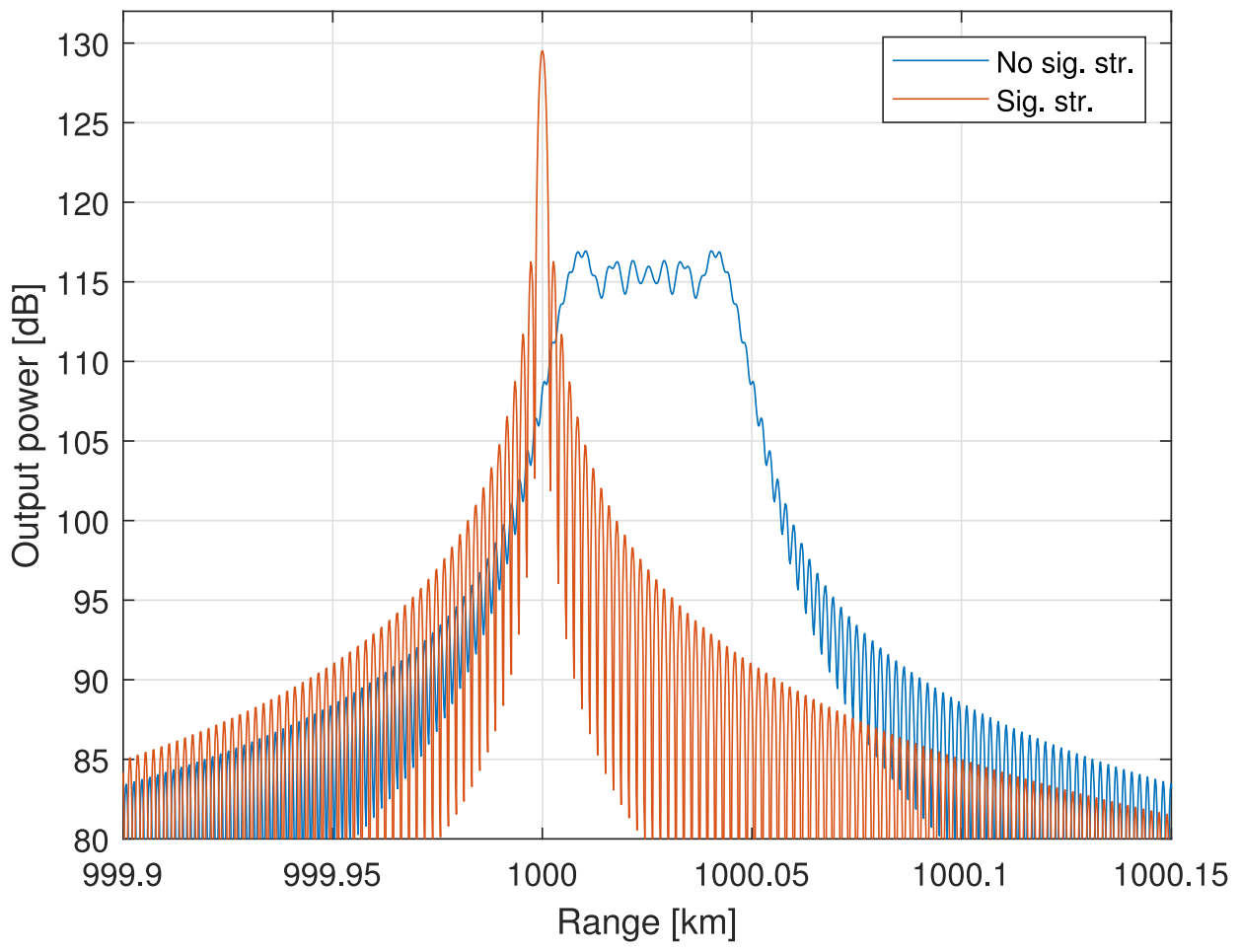

**FIGURE 4** | Correlation of a simulated noise-free target with the CnG without (blue) and with (red) signal stretch (intra-pulse range migration). The waveform is a chirp and the target and system parameters are given in Table 1, Sim. II.

## 5.4 | Comparison to Other Methods

A direct one-to-one comparison with existing RDA approaches is not straightforward. To the best of the authors' knowledge, no previously published RDA method explicitly compensates intra-pulse baseband stretch. Consequently, existing techniques can only be compared under partial equivalence, typically by augmenting them with an additional acceleration hypothesis bank where required.

The computational complexity is an important aspect of performance. Let $N_{\mathrm{a}}, N_{\mathrm{v}}$, and $N_{\mathrm{r}}$ denote the number of radial acceleration, radial velocity, and range samples, respectively. Let $N_{\mathrm{t}}$ be the number of samples required to represent the transmitted pulse in baseband, and let $N_{\mathrm{p}}$ be the number of pulses. We consider a case where $N_{\mathrm{v}}$ could be larger than $N_{\mathrm{p}}$, and $N_{\mathrm{r}}$ is in the same order of $N_{\mathrm{t}}$.

A brute-force Time-Domain (TD)-MF would then require a computational complexity of $\mathcal{O}(N_{\mathrm{a}}N_{\mathrm{v}}N_{\mathrm{p}}N_{\mathrm{t}}^2)$. In contrast, the CnG method requires, for each $(v_0, a_0)$ hypothesis and for each pulse, a FFT of length on the order of $N_{\mathrm{t}}$. This is the core of our method, being able to apply FFT in fast time while compensating for pulse stretch. Since the upsampling is done once per CPI, a linear interpolation is sufficient. Therefore, the interpolation load is additive and doesn't impact the overall complexity, which is $\mathcal{O}(N_{\mathrm{a}}N_{\mathrm{v}}N_{\mathrm{p}}N_{\mathrm{t}}\log N_{\mathrm{t}})$. This constitutes a substantial reduction compared to TD-MF, particularly for long pulses where $N_{\mathrm{t}}$ is large (as commonly encountered in space surveillance systems).

Most of the other approaches have a lower computational complexity which comes at various trade-offs. The beat-frequency of the LFM allows a single fast-time FFT, without the need of a subsequent IFFT. Table 2 attempts to compare these methods in a clear manner. The CnG is the only method developed to accommodate any waveform structure, intra-pulse motion and intra-pulse baseband stretch. It potentially allows the use of non-LFM, thus avoiding the RD-coupling ambiguity altogether.

## 6 | Method Generalisation

An analysis of the limitations of the CnG method is presented, in order to quantify for which systems/scenarios it is suitable. The residual error arises from uncorrected quadratic motion within each pulse. As a result, under conditions of high radial acceleration and long pulse durations, the method may still exhibit performance degradation.

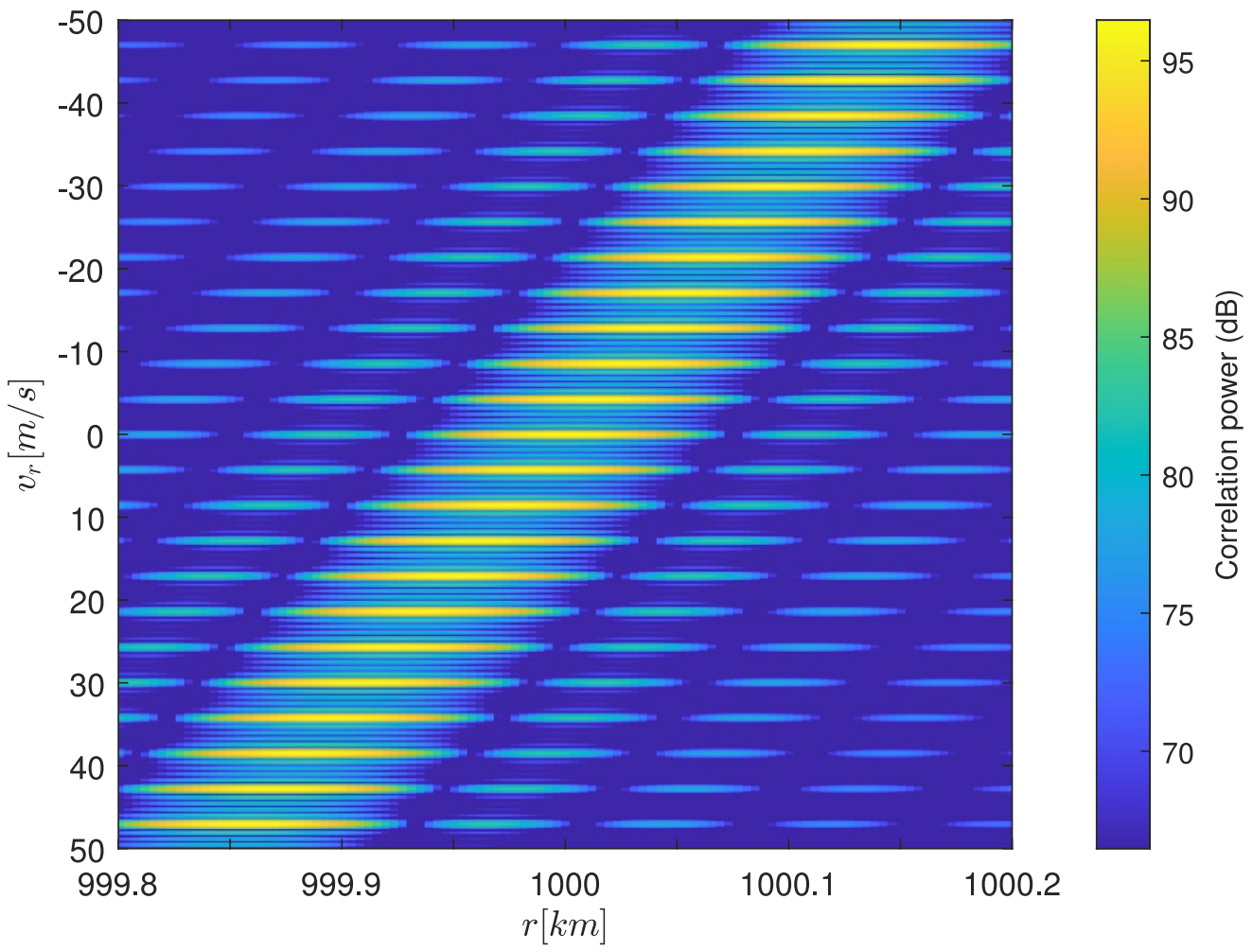

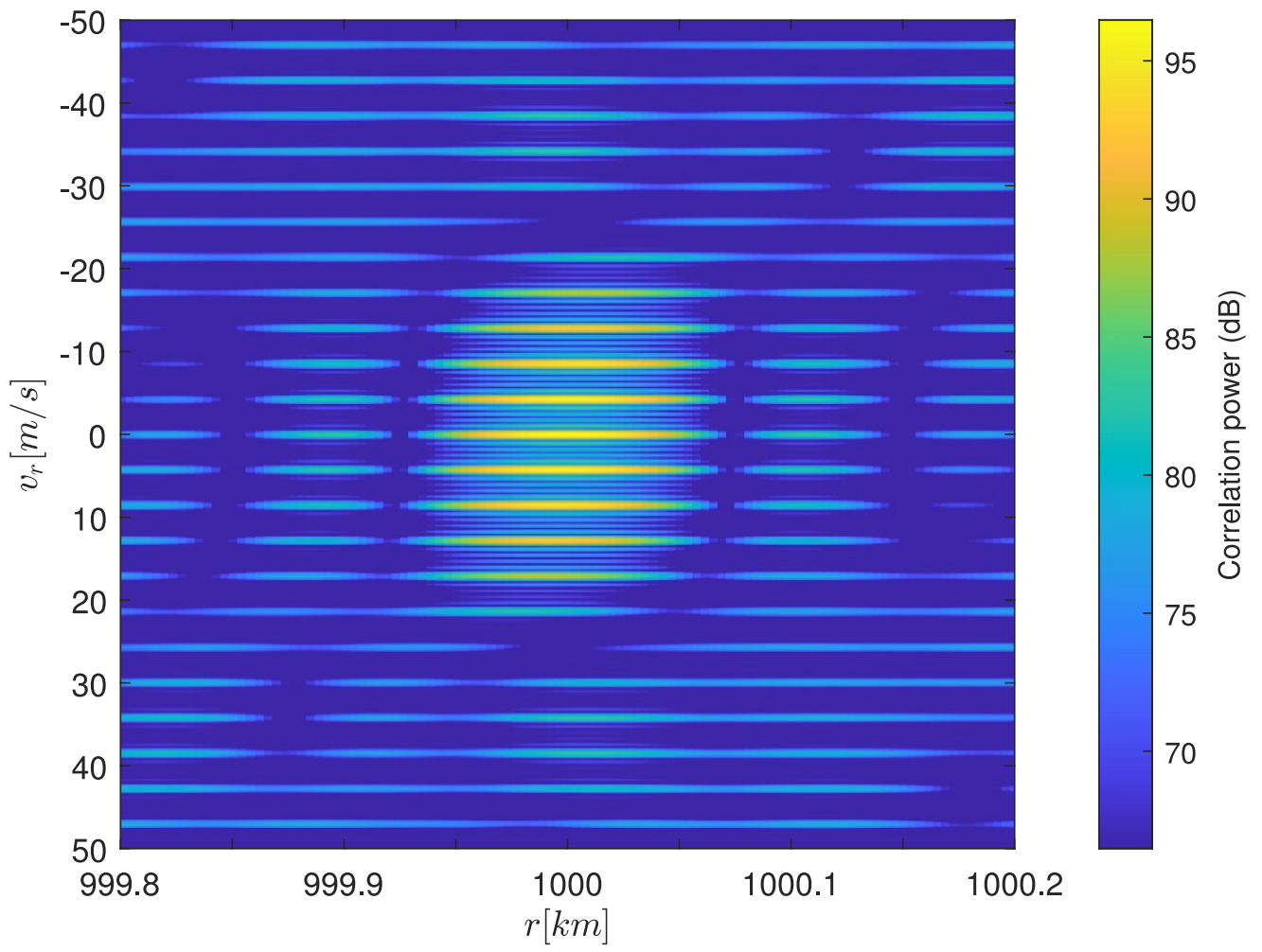

**FIGURE 5** | RD maps with LFM (left) and Costas (right) waveform for a simulated target. The target and system parameters are given in Table 1, Sim. III.

**TABLE 2** | Comparison of RDA processing approaches.

| Method | Comments | Computational complexity | Intra-pulse Doppler shift | Intra-pulse baseband stretch |
|---|---|---|---|---|
| TD-MF | Exact motion model | $\mathcal{O}(N_a N_v N_p N_t^2)$ | ✓ | ✓ |
| **CnG** | Proposed method in this manuscript | $\mathcal{O}(N_a N_v N_p N_t \log N_t)$ | ✓ | ✓ |
| Higher-order KT + Acc. bank | Doppler ambiguity challenges | $\mathcal{O}(N_a N_v N_t \log(N_p))$ | ✗ | ✗ |
| RFRFT + Acc. bank | Could be extended to compensate intra-pulse Doppler shift | $\mathcal{O}(N_a N_v N_p N_t \log(N_p))$ | ✗ | ✗ |
| DPT | Requires high SNR | $\mathcal{O}(N_t \log(N_t) N_p \log(N_p))$ | ✓ | ✗ |
| Coupling-based LFM method | Resolves LFM RD-coupling ambiguity | $\mathcal{O}(N_a N_v N_t \log(N_t N_p))$ | ✓ | ✗ |

## 6.1 | Unified Metric

To analyse the possible residual error of the CnG assumption, a noise-free target was simulated across a wide range of parameter combinations, listed in Table 3. For all those 1664 parameter combinations listed there, Equation (25) was computed and compared with the theoretical accurate value from Equation (5). We refer to the ratio between them as the correlation loss, expressed in decibels. It was empirically found that this loss can be accurately predicted using a specific dimensionless ratio that depends on both system and scene parameters. The so-called *acceleration ratio* is

$$\Upsilon = a T_p (T_p + \tau)/\lambda_c. \tag{26}$$

Similar acceleration-induced coherence limits have been derived repeatedly in different radar applications (e.g., [1, 16]).

Figure 6 shows the observed correlation loss as a function of $\Upsilon$, alongside the $\mathrm{sinc}^2$ function for comparison. The close alignment of nearly all data points with the $\mathrm{sinc}^2$ curve-despite the wide range of parameter variations-suggests that this function serves as a reliable estimator of the correlation loss. We now obtain a practical tool for the radar system designer, providing a simple means to estimate the expected performance of the CnG method. While the parameters $a, \tau$ are determined by the observed scene, namely the target's radial acceleration and range, the radar designer retains control over key system parameters such as the pulse duration $T_p$ and the carrier wavelength $\lambda_c$.

Having examined the effect of specific values of the acceleration ratio $\Upsilon$, we next consider the physical scenarios in which large values of $\Upsilon$ may realistically occur and lead to a significant SNR degradation. From Equation (26), it is evident that $T_p$ enters quadratically, indicating that long pulse durations can significantly increase $\Upsilon$. In many radar configurations, particularly monostatic systems where simultaneous transmission and reception is not possible, the round-trip delay $\tau$ is often of the same order of magnitude as the pulse length. As a consequence of this quadratic dependence, spaceborne objects represent the most likely class of targets to yield large values of $\Upsilon$.

As an illustrative example, $\Upsilon$ has been evaluated for a ground-based radar located on the equator observing objects on circular equatorial orbits. Both prograde orbits, corresponding to motion in the same direction as Earth's rotation, and retrograde orbits, corresponding to motion in the opposite direction, are considered over a range of orbital heights. The values of $\tau$ and $a$ are evaluated at the instant when the target passes through zenith above the radar, which typically corresponds to the point of maximum radial acceleration. The pulse duration is chosen as $T_p = \tau$, and a carrier frequency of 10 GHz is assumed (X-band, widely used in spaceborne and ground-based radar systems), corresponding to a wavelength of $\lambda_c = 3\,\text{cm}$. The resulting values of $\Upsilon$ are shown in Figure 7.

**TABLE 3** | General simulation parameters.

| Symb. | Parameter | Value |
|---|---|---|
| $f_c$ | Carrier frequency | 650 MHz, 1300 MHz |
| $B$ | Bandwidth | 2 MHz |
| $T_{pr}$ | Pulse rep. Interval | 25 ms, 50 ms, 75 ms, 100 ms |
| $T_p$ | Pulse duration | 4 ms, 8 ms |
| $r_0$ | Range of target | 3000 km, 10000 km |
| $v_0$ | rad. vel. of target | 0, 4000 $ms^{-1}$ |
| $a_0$ | rad. acc. of target | 0, 50 $ms^{-2}$, ..., 600 $ms^{-2}$ |
| $N_p$ | No. of pulses | 12, 24 |

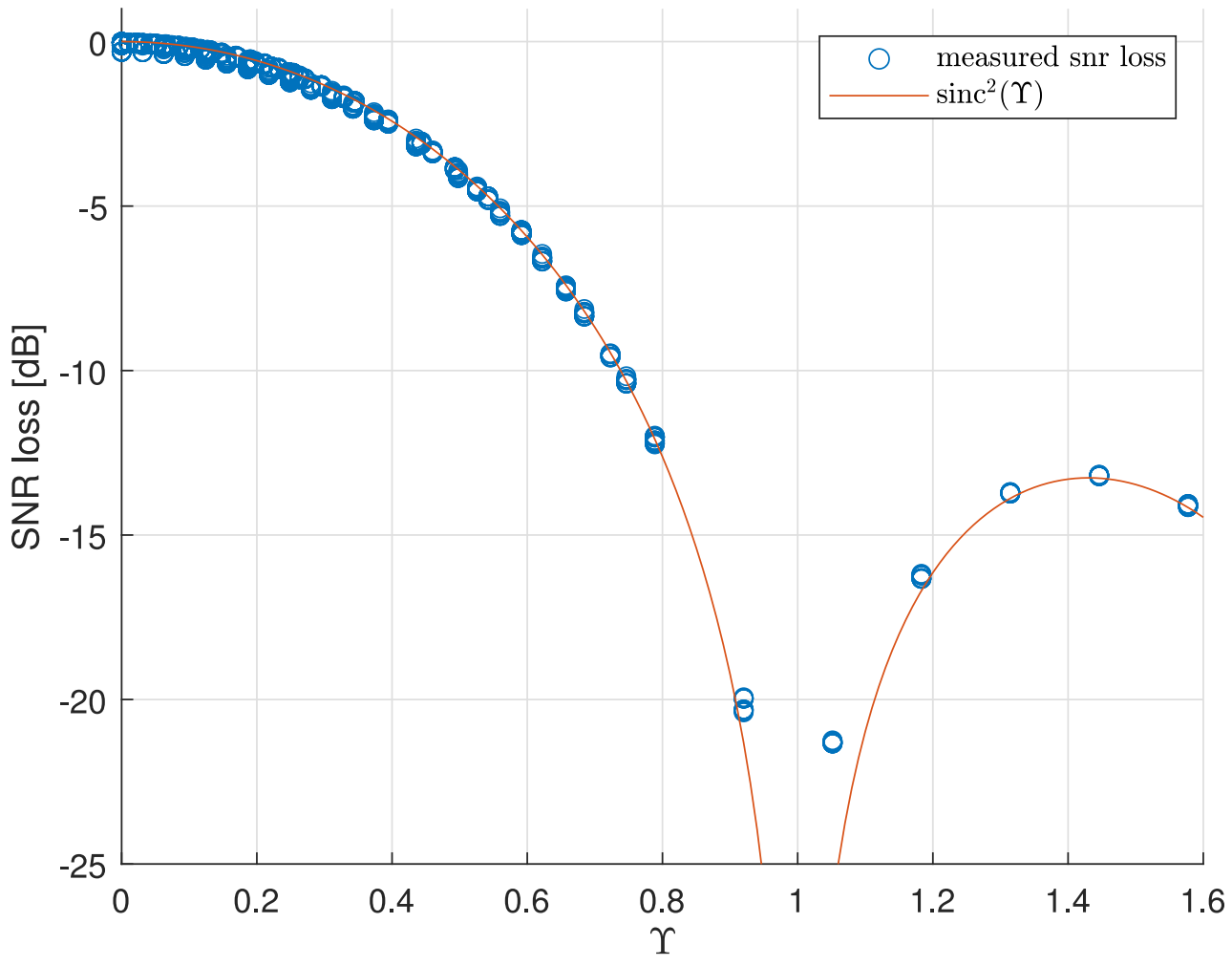

**FIGURE 6** | SNR loss of CnG method in dependence on acceleration ratio. In every case, a target was simulated and processed at the correct $v_0$ and $a_0$ and $r_0$. The resulting power, divided by the theoretical power of a perfect signal compensation, was plotted here, against the acceleration ratio. The parameter combinations used are given in Table 3.

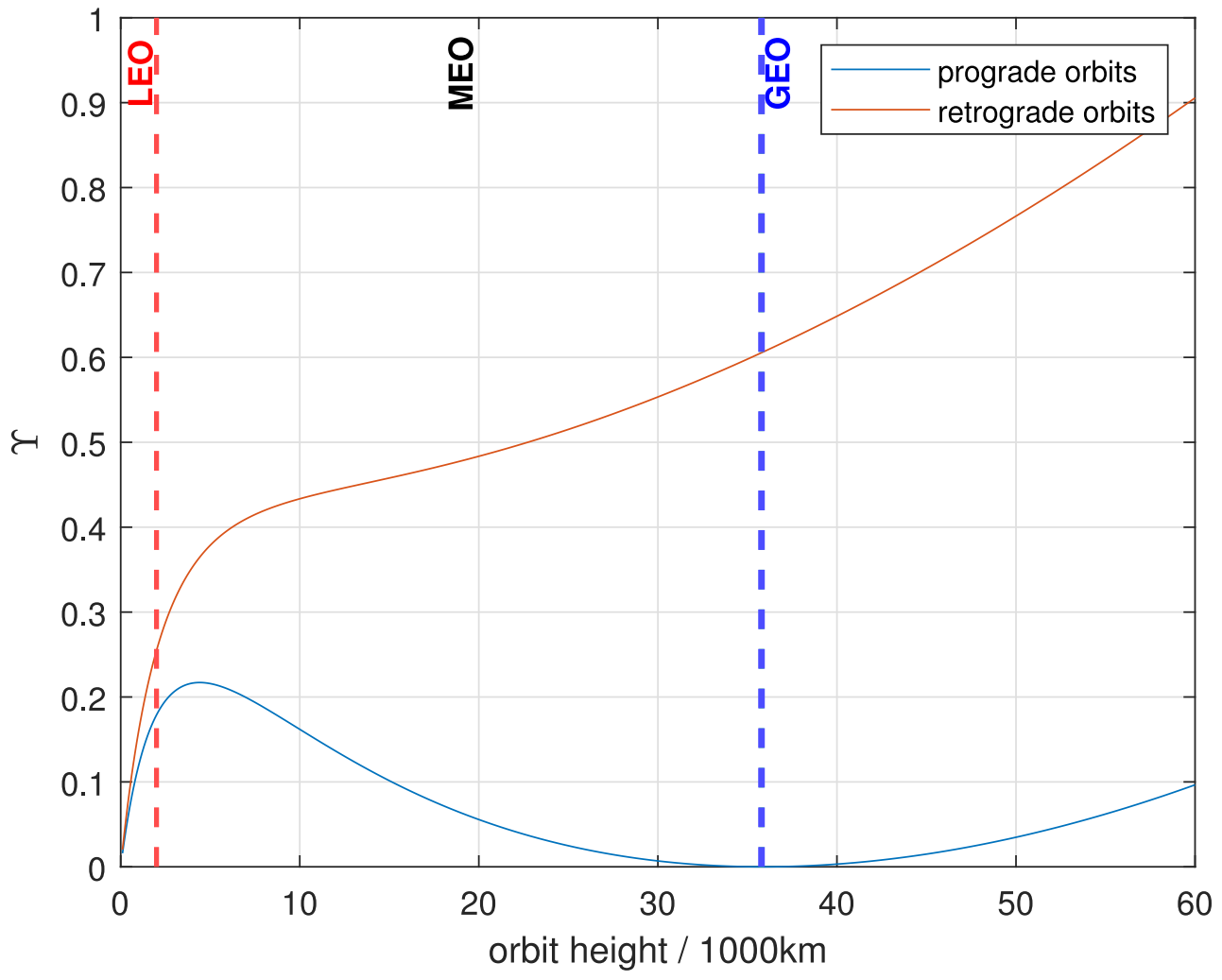

**FIGURE 7** | $\Upsilon$ values for different space objects on equatorial orbits (prograde and retrograde) passing directly over a radar at equator. Radar parameters: $T_p = \tau$, $\lambda_c = 3\,\text{cm}$.

The plot indicates that significant values of $\Upsilon$ occur primarily for high-altitude retrograde orbits. For prograde orbits, $\Upsilon$ approaches zero at the Geostationary Orbit (GEO) altitude (where radial acceleration is zero). It should also be noted that the overall magnitude of $\Upsilon$ scales inversely with the carrier wavelength $\lambda_c$. Over 98 % of operational satellites are at or below GEO altitudes [17], where $\Upsilon < 0.5$ and the overall SNR loss is smaller than 3 dB. This indicates that the CnG method is applicable for space surveillance scenarios. For orbits outside the equatorial plane, it is assumed that the resulting curve will lie somewhere between the prograde and retrograde orbits, depending on the inclination.

## 6.2 | Statistical Noise Analysis

In addition to the deterministic motion-compensation aspects discussed above, it is important to assess the statistical estimation performance of the proposed CnG method under additive noise. In this subsection, we compare the empirical Root Mean Square Error (RMSE) of the joint $(r_0, v_0, a_0)$ estimate obtained by Monte-Carlo (MC) simulations to the corresponding Cramér-Rao-Bound (CRB). The simulation was done for three different sets of parameters as detailed in Table 4.

It was found that the results were almost identical for the three parameter sets, hence only the result of the parameter set III is presented. For this fixed scenario, the MC simulation was conducted over a range of input SNR values. For each SNR level, 643 independent realisations of white Gaussian noise were generated and added to the signal, the full three-dimensional RDA map was computed using the proposed method, and the target parameters were estimated from the location of the global maximum. The maximum was obtained by sampling on a fine grid, followed by a tricubic interpolation.

The CRB was subsequently computed (see [18]) for the specific parameter set. It is important to note, that the CRB was determined for the actual data model according to Equations (2) and (4) without the CnG approximation. The estimator on the other hand uses the approximation. It is therefore expected that the estimator, although being effectively a maximum-likelihood estimator, under the assumption of a single target and when the true data follows the CnG approximation, might not reach the CRB. It is further expected that a bias, denoting the ultimate mismatch of the CnG approximation, may occur (see Appendix for more derivation details).

**TABLE 4** | Statistical simulations parameters values.

| Symb. | Parameter | Sim. I | Sim. II | Sim. III |
|---|---|---|---|---|
| $f_c$ | Carrier frequency | 1.3 GHz | 1.3 GHz | 1.3 GHz |
| $B$ | Bandwidth | 2 MHz | 2 MHz | 2 MHz |
| $T_{\text{pr}}$ | PRI | 30 ms | 30 ms | 30 ms |
| $T_{\text{p}}$ | Pulse duration | 4.5 ms | 4.5 ms | 4.5 ms |
| $r_0$ | Range of target | 1000 km | 1000 km | 1000 km |
| $v_0$ | rad. vel. of target | 0 $\text{ms}^{-1}$ | 4000 $\text{ms}^{-1}$ | −4000 $\text{ms}^{-1}$ |
| $a_0$ | Rad. acc. of target | 0 $\text{ms}^{-2}$ | 250 $\text{ms}^{-2}$ | 250 $\text{ms}^{-2}$ |
| $N_{\text{p}}$ | No. of pulses | 12 | 12 | 12 |

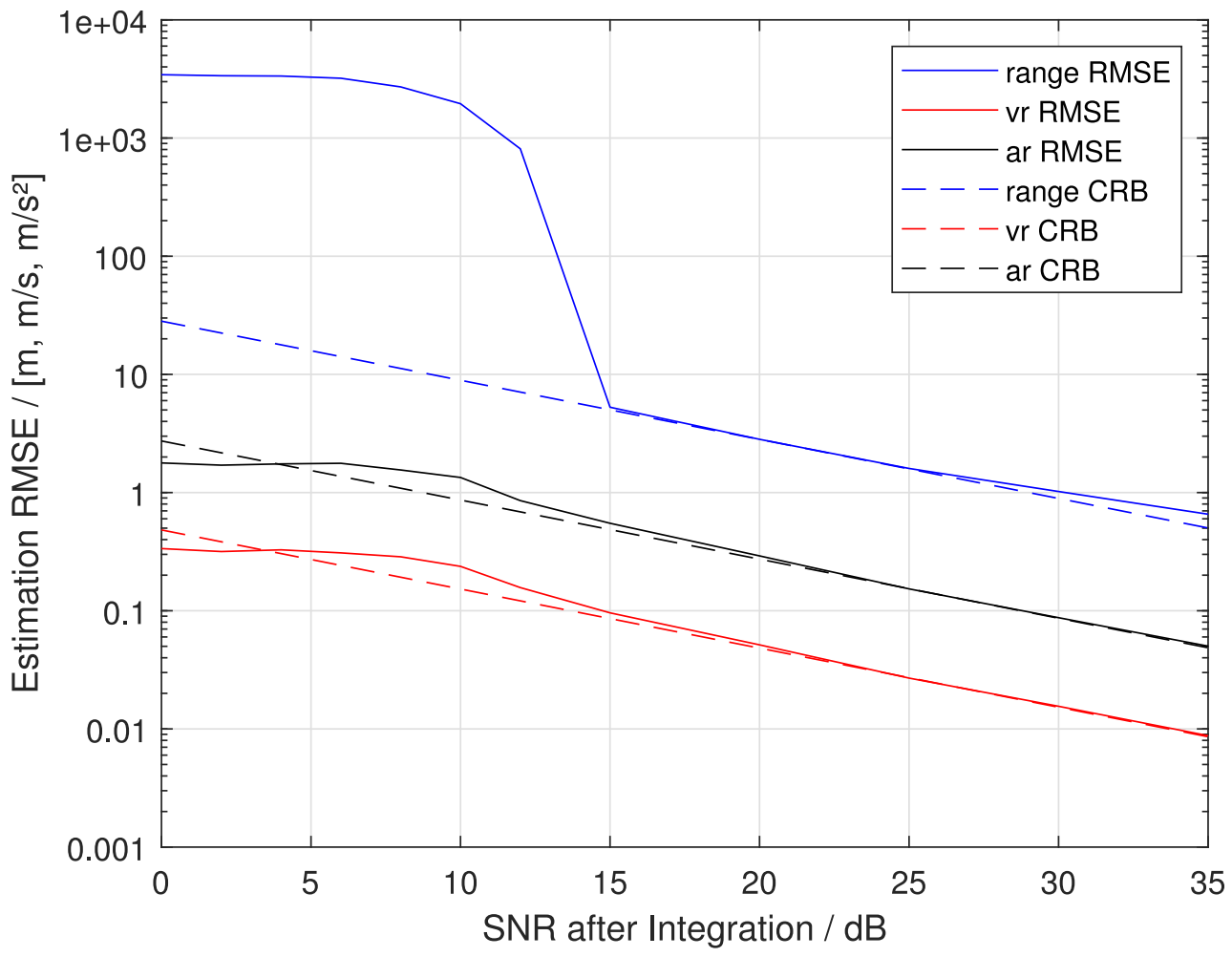

**FIGURE 8** | RMSE and CRB of different variables in a logarithmic plot. Simulation parameters are given in Table 4, Sim. III.

Figure 8 shows the RMSE of the three motion parameters and the respective CRB, that is the root diagonal element of the inverse Fisher Information Matrix (FIM). For SNR values below ≈15 dB a large discrepancy between the RMSE and the CRB can be seen. Overall this resembles the expected behaviour of a maximum-likelihood estimator within its threshold region. The shape of the RMSE curves for the radial velocity and acceleration are misleading in that region as they indicate a variance of less than the CRB. This is, however, solely due to the implementation of the estimator. The estimation for both parameters is performed using a grid based search where the minimum and maximum values of the grid are limited for performance reasons. For the low SNR cases the true variance of the estimator indeed exceeds the width of the search grid and the resulting RMSE curve is representing the grid rather than the estimation error. As the SNR increases the error becomes well confined in the search grid and thus the empirical RMSE of the simulation resembles the true RMSE of the estimator. Starting at ≈15 dB the RMSE approaches the CRB quite well, showing again the behaviour known from maximum likelihood estimators.

For very high SNR values and nonzero acceleration, the range error diverges from the CRB due to a small bias (originating from the CnG approximation). It can overall be said that the CRB is attained.

## 7 | Conclusions

Although the proposed method has a higher computational complexity than existing approaches, it is, to the best of the authors' knowledge, the only one that explicitly accounts for intra-pulse baseband stretch. Simulation results demonstrated accurate refocusing of target responses under strong acceleration. To characterise the limits of the underlying CnG approximation, a dimensionless acceleration ratio was introduced, linking scene dynamics and radar parameters to predictable correlation loss. MC simulations further showed that the estimator approaches the CRB, indicating that performance is primarily limited by noise rather than by model mismatch. Overall, the proposed framework provides a practical and computationally feasible solution for RDA estimation in FMAT scenarios, such as SSA, where both intra-pulse Doppler shift and baseband stretch, as well as quadratic motion across pulses cannot be neglected.

### Author Contributions

**Nadav Neuberger:** conceptualization, validation, visualization, writing – original draft, writing – review and editing. **Simon Kollecker:** validation, visualization, writing – original draft, writing – review and editing. **Martin Käske:** writing – review and editing.

### Acknowledgements

Open Access funding enabled and organized by Projekt DEAL.

### Funding

The authors have nothing to report.

### Conflicts of Interest

The authors declare no conflicts of interest.

### Data Availability Statement

Data sharing not applicable to this article as no datasets were generated or analysed during the current study.

## Appendix

### Approximations Analysis

### CnG Limits Analysis

Presented here is a theoretical analysis of the assumption made in the new approach. It provides a sufficient condition for when the assumption is justified. The goal is to quantify the corresponding inaccuracies as a function of target and system parameters.

For $t \in [0, T_{\text{cp}}]$, let $r_{\max} = \max_t r(t)$, $\tau_{\max} = \max_t \tau(t)$ and $v_{\max} = \max_t |\dot{r}(t)|$. Note the absolute value in the definition of $v_{\max}$. The only approximation made here is the linearisation of the motion inside the single pulse. The general formula of the delay is (1), whereas the linearised version is (18), so we are trying to justify

$$2\frac{r(t-\tau(t)/2)}{c_0} \approx \frac{2r_m + 2v_m\Delta t}{c_0 + v_m}. \tag{A1}$$

To this end, let us define the terms

$$\begin{aligned}
\varepsilon_1(t) &= \frac{a_0\tau^2(t)}{4(c_0+v_0+a_0t)}\\
\varepsilon_2(t) &= -\frac{a_0\Delta t\left(2r_0+2v_0t+a_0t^2\right)}{(c_0+v_0+a_0T_m)(c_0+v_0+a_0t)}\\
\varepsilon_3(t) &= -\frac{a_0\Delta t^2}{c_0+v_0+a_0t}\\
\varepsilon_\Sigma(t) &= \varepsilon_1(t)+\varepsilon_2(t)+\varepsilon_3(t).
\end{aligned}$$

These are the neglected terms, because

$$\begin{aligned}
\tau(t,\theta) &= \frac{2r\left(t-\frac{\tau(t)}{2}\right)}{c_0}\\
&= \frac{2r_0+2v_0t+a_0t^2+\frac{1}{4}a_0\tau^2(t)}{c_0+v_0+a_0t}\\
&= \frac{2r_0+2v_0t+a_0t^2}{c_0+v_0+a_0t}+\varepsilon_1(t)\\
&= \frac{2r_0+2v_0t+a_0t^2}{c_0+v_0+a_0T_m}+\varepsilon_1(t)+\varepsilon_2(t)\\
&= \frac{2r_0+2v_0t+a_0t^2-a_0\Delta t^2}{c_0+v_0+a_0T_m}+\varepsilon_1(t)+\varepsilon_2(t)+\varepsilon_3(t)\\
&= \frac{2r_0+2v_0T_m+a_0T_m^2+2v_0\Delta t+2a_0T_m\Delta t}{c_0+v_0+a_0T_m}+\varepsilon_\Sigma(t)\\
&= \frac{2r_m+2v_m\Delta t}{c_0+v_m}+\varepsilon_\Sigma(t),
\end{aligned}$$

where the fourth equality comes from the fact that

$$\frac{a}{b+\varepsilon} = \frac{a}{b} - \frac{a\varepsilon}{b(b+\varepsilon)}. \tag{A2}$$

In order to limit $\varepsilon$ values we observe that $0 \le \Delta t \le T_{\text{p}} + \tau_{\max}$ is the time interval used in the computation. Therefore, the errors are bound by

$$\begin{aligned}
|\varepsilon_1(t)| &\le \frac{|a_0|\tau_{\max}^2}{4(c_0-v_{\max})}\\
|\varepsilon_2(t)| &\le \frac{|a_0|\left(T_{\text{p}}+\tau_{\max}\right)r_{\max}}{(c_0-v_{\max})^2}\\
|\varepsilon_3(t)| &\le \frac{|a_0|\left(T_{\text{p}}+\tau_{\max}\right)^2}{(c_0-v_{\max})^2}.
\end{aligned}$$

Hence, a sufficient condition is that all of the terms above are significantly lower than $1/f_c$.

### CRB Derivatives

For the CRB calculation, the derivatives of the signal model (2) with respect to the target parameters are needed. The complex target amplitude $A$, although being ignored so far, is an unknown in terms of the CRB that needs to be considered. The five estimation variables are: $r_0, v_r, a_r, \Re(A), \Im(A)$ with $\Re(...)$ and $\Im(...)$ being the real and imaginary part respectively. The derivative of (2) can be computed using the implicit formula.

For the computation of the derivative of the signal model (2), the derivative of the implicit formula for $\tau$ is needed. The formula for the derivative will hence also be implicit and also have an explicit dependence on $\tau$. If, however, $\tau(t)$ is computed before (using either the analytical solution or a fixed point iteration to resolve the implicit part), then the dependence of the derivative of $\tau$ on itself is a linear one and can be trivially resolved by rearranging, for example

$$\frac{\partial \tau}{\partial v_r} = \frac{2}{c_0}(t - \tau(t)/2) + \frac{2}{c_0}\dot{r}(t - \tau(t)/2)\cdot(-0.5)\frac{\partial \tau}{\partial v_r}$$
$$\Leftrightarrow \frac{\partial \tau}{\partial v_r} = \frac{2t - \tau(t)}{c_0 + v_r + a_r(t - \tau(t)/2)}.$$

The computation of the FIM can be performed using the statement from the beginning of [18, Chapter 3.9], where the derivative of the signal $s$ itself can be computed via the Fourier domain. The inverse of the FIM constitutes a lower bound on the covariance matrix of the estimation parameters.